%% file: eccv2020a.tex
\documentclass[runningheads]{llncs}
\usepackage{graphicx}
\usepackage{amsmath,amssymb} % define this before the line numbering.

\newcounter{submission}
\ifnum\value{submission}=1
\usepackage{ruler}
\usepackage[width=122mm,left=12mm,paperwidth=146mm,height=193mm,top=12mm,paperheight=217mm]{geometry}
\fi
\usepackage{color}

\usepackage[square,numbers,sort&compress]{natbib}
\usepackage{booktabs}
\usepackage{paralist}
\usepackage{xcolor}
\usepackage{hyperref}
\definecolor{darkred}{rgb}{0.85,0,0}
\definecolor{darkgreen}{rgb}{0,0.6,0}
\definecolor{darkblue}{rgb}{0,0,0.85}
\hypersetup{hidelinks, breaklinks, colorlinks,
  linkcolor={darkred}, citecolor={darkgreen}, urlcolor={darkblue}}
\usepackage{url}
\graphicspath{{images/}}
\DeclareGraphicsExtensions{.pdf,.png,.jpg}

\newcommand{\ie}{\emph{i.e.},}
\newcommand{\eg}{\emph{e.g.},}

\newcommand{\etal}{\emph{et~al.}}
\newsavebox{\spaceofwidthbox}
\newcommand{\spaceofwidth}[1]%
{\savebox{\spaceofwidthbox}{#1}\hspace*{\wd\spaceofwidthbox}}
\newcommand{\sig}{$^{*}$}
\newcommand{\notsig}{\spaceofwidth{$^{*}$}}

\begin{document}
% \renewcommand\thelinenumber{\color[rgb]{0.2,0.5,0.8}\normalfont\sffamily\scriptsize\arabic{linenumber}\color[rgb]{0,0,0}}
% \renewcommand\makeLineNumber {\hss\thelinenumber\ \hspace{6mm} \rlap{\hskip\textwidth\ \hspace{6.5mm}\thelinenumber}}
% \linenumbers
\pagestyle{headings}
\mainmatter
\def\ECCVSubNumber{1966}  % Insert your submission number here

\title{Object classification from randomized EEG trials}

\ifnum\value{submission}=1
\titlerunning{ECCV-20 submission ID \ECCVSubNumber}
\authorrunning{ECCV-20 submission ID \ECCVSubNumber}
\author{Anonymous ECCV submission}
\institute{Paper ID \ECCVSubNumber}
\else
\titlerunning{Object classification from randomized EEG trials}
\authorrunning{H. Ahmed \etal}
\author{Hamad Ahmed\inst{1}\orcidID{0000-0002-9524-4467} \and
  Ronnie B Wilbur\inst{1}\orcidID{0000-0001-7081-9351} \and
  Hari M Bharadwaj\inst{1}\orcidID{0000-0001-8685-9630} \and
  Jeffrey Mark Siskind\inst{1}\orcidID{0000-0002-0105-6503}}
\institute{Purdue University, West Lafayette IN 47907-2035, USA
\email{\{ahmed90,wilbur,hbharadw,qobi\}@purdue.edu}}
\fi

\maketitle

\begin{abstract}
  New results suggest strong limits to the feasibility of classifying human
  brain activity evoked from image stimuli, as measured through EEG.
  Considerable prior work suffers from a confound between the stimulus class
  and the time since the start of the experiment.
  A prior attempt to avoid this confound using randomized trials was unable to
  achieve results above chance in a statistically significant fashion when the
  data sets were of the same size as the original experiments.
  Here, we again attempt to replicate these experiments with randomized trials
  on a far larger (20$\times$) dataset of 1,000 stimulus presentations of each
  of forty classes, all from a single subject.
  To our knowledge, this is the largest such EEG data collection effort from a
  single subject and is at the bounds of feasibility.
  We obtain classification accuracy that is marginally above chance and above
  chance in a statistically significant fashion, and further assess how
  accuracy depends on the classifier used, the amount of training data used,
  and the number of classes.
  Reaching the limits of data collection without substantial improvement in
  classification accuracy suggests limits to the feasibility of this
  enterprise.
\end{abstract}

\section{Introduction}

\begin{figure}[t]
  \centering
  \resizebox{\textwidth}{!}{\begin{tabular}{@{}rl@{}}
      \raisebox{18pt}{(a)}&\includegraphics[height=72pt]{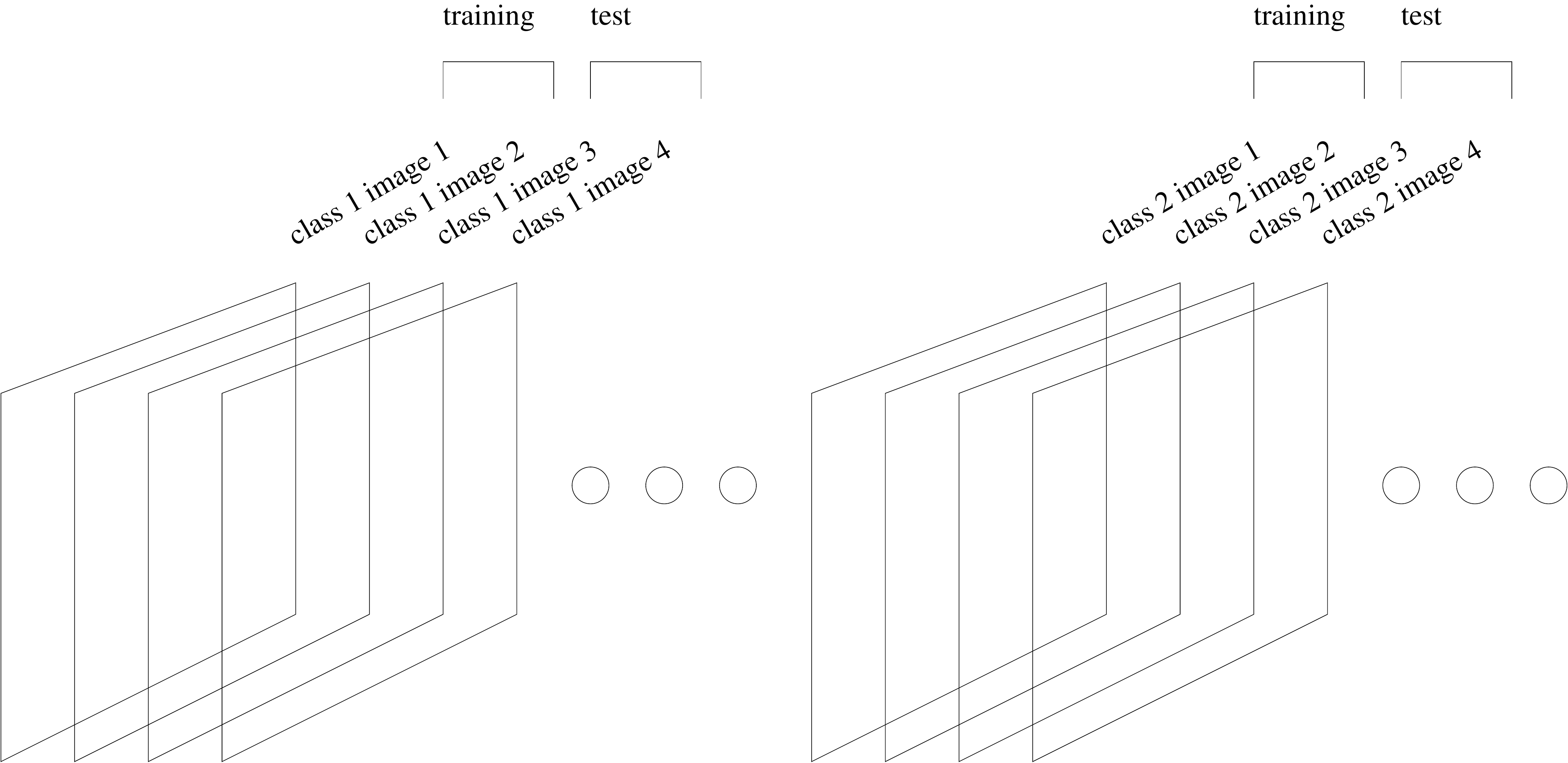}\\
      \raisebox{18pt}{(b)}&\includegraphics[height=72pt]{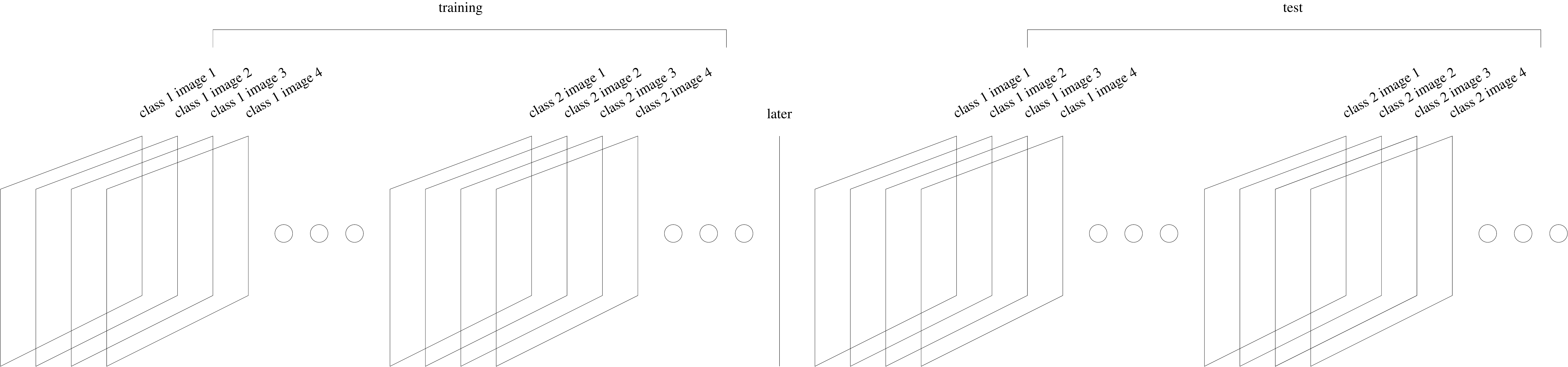}\\
      \raisebox{18pt}{(c)}&\includegraphics[height=72pt]{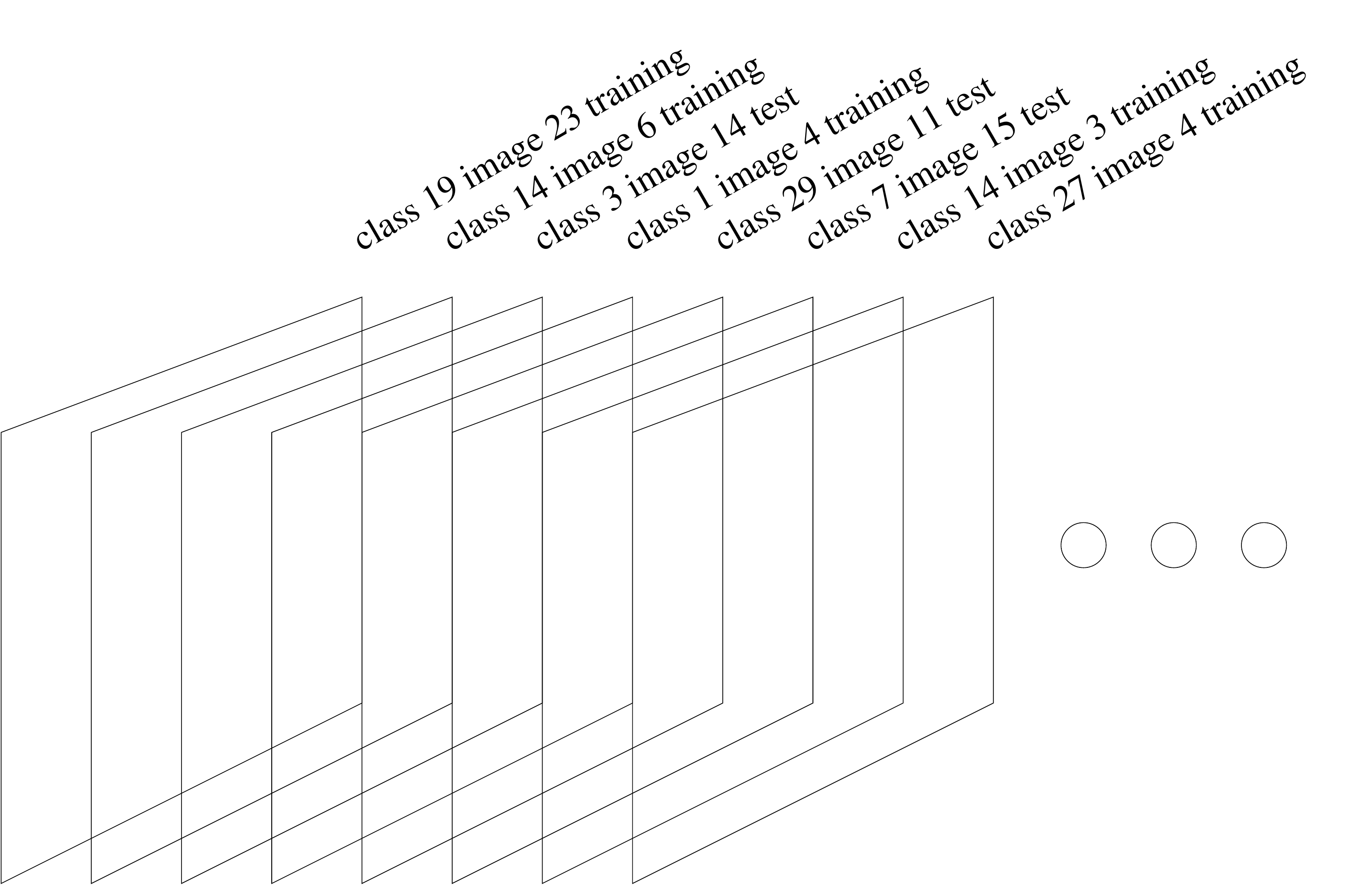}\\
  \end{tabular}}
  \caption{Stimulus presentation order and training/test splits employed by
    (a)~Spampinato \etal\citep{spampinato2017},
    (b)~Cudlenco \etal\citep{Cudlenco2019}, and (c)~randomized trials.
    (a) and (b) confound stimulus class with time since beginning of
    experiment.}
  \label{fig:designs}
\end{figure}

There has been considerable recent interest in applying deep learning to
electroencephalography (EEG).\@
% 123 249
Two recent survey papers \citep{Craik2019, roy2019} collectively contain 372
references.
Much of this work attempts to classify human brain activity evoked from visual
stimuli.
A recent CVPR oral \citep{spampinato2017} claims to decode one of forty object
classes when subjects view images from ImageNet \citep{deng2009} with 82.9\%
accuracy.
Considerable follow-on work uses the same dataset \citep{palazzo2017,
  kavasidis2017, palazzo2018, palazzo2018recent, du2018multi, Bozal2017,
  jiang2019context, jiaodecoding, zhang2018using, du2019doubly, zhong2018elstm,
  mukherjee2019cogni, fares2018region, hwang2019ezsl}.
Our recent paper \citep{li2020} demonstrates that this classification accuracy
is severely overinflated due to flawed experiment design.
All stimuli of the same class were presented to subjects as a single block
(Fig.~\ref{fig:designs}a).
Further, training and test samples were taken from the same block.
Because all EEG data contain long-term temporal correlations that are unrelated
to stimulus processing and their design confounded block-effects with class
label, Spampinato \etal\citep{spampinato2017} were classifying these long-term
temporal patterns, not the stimulus class.
Because the training and test samples were taken in close temporal proximity
from the same block, the temporal correlations in the EEG introduced label
leakage between the training and test data sets.
When the experiment of Spampinato \etal\citep{spampinato2017} is repeated with
randomized trials, where stimuli of different classes are randomly intermixed,
classification accuracy drops to chance \citep{li2020}.

Another recent paper \citep{Cudlenco2019} attempts to remedy the shortcomings
of a block design by recording two different sessions for the same subject, each
organized as a block design, one to be used as training data and one to be used
as test data.
However, both sessions used the same stimulus presentation order
(Fig.~\ref{fig:designs}b).
Our recent paper \citep{li2020} demonstrates that classification accuracy can
even be severely inflated with such a cross-session design that employs the
same stimulus presentation order in both sessions due to the same long-term
transients that are unrelated to stimulus processing.
While an analysis of training and test sets coming from different sessions with
the same stimulus presentation order yields lower accuracy than an analysis
where they come from the same session, accuracy drops to chance when the two
sessions have different stimulus presentation order.

All this prior work is fundamentally flawed due to improper experiment design.
Essentially, the EEG signal encodes a clock and any experiment design where
stimulus class correlates with time since beginning of experiment allows
classifying the clock instead of the stimuli.
This means that all data collected in this fashion is irreparably contaminated.

% \ added to prevent overfull hbox
We previously \citep{li2020} attempted to replicate the experiment of Spampinato
\etal\ \citep{spampinato2017} six times with nine different classifiers,
including the LSTM employed by them, with randomized trials
(Fig.~\ref{fig:designs}c) instead of a block design.
All attempts failed, yielding chance performance.

%\needswork: cite other image classification EEG experiments that use a block
%            design like tirupattur2018 and its origin, list from survey paper

% post tpami2019 revision to arxiv and update citation
% stress that all prior work using block design is invalid and the reported
% accuracies are wildely inflated, they are not known to outperform chance
% state that this is the only known EEG classification task with more than N
%   classes that achieves accuracy above chance
% look at Spampinatro nonimage/object [5, 14, 1, 23]
%                     image/object [26, 4, 20, 2, 11, 22, 10]
% update CNN image: 409, 5
% regressor idea

Here, we ask the following four questions:
\begin{compactenum}
\item\emph{Is it possible to decode object class from EEG data recorded from
  subjects viewing image stimuli with randomized stimulus presentation order?}
\item\emph{If so, how many distinct classes can one decode?}
\item\emph{If so, how much training data is needed?}
\item\emph{If so, what classification architectures allow such decoding?}
\end{compactenum}
To answer these questions, we collected EEG recordings from 40,000 stimulus
presentations to a single subject.
To our knowledge, this is by far the largest recording effort of its kind.
Moreover, we argue that collecting such a large corpus is at the bounds of
feasibility; it is infeasible to collect any appreciably larger corpus.
With this corpus we achieve modest ability to decode stimulus class with
accuracy above chance in a statistically significant fashion.
By using a greedy method to determine the most discriminable $n$~classes for
$2\leq n\leq 40$, and determining the classification accuracy for each such
set, we show that forty classes is at the limit of feasibility.
Further, by repeating the experiments with successively larger fractions of the
dataset, we determine that at least half of this large dataset is needed to
achieve this accuracy.
Finally, we show that an LSTM architecture previously reported to yield high
accuracy \citep{spampinato2017} is unable to achieve classification accuracy
above chance in a statistically significant fashion.
The only two classifiers that we tried that achieve classification accuracy
above chance in a statistically significant fashion are a support vector
machine (SVM \citep{cortes1995}) and the one-dimension convolutional neural
network (1D~CNN) previously reported by us \citep{li2020}.

\section{Data Collection}

Spampinato \etal\citep{spampinato2017} selected fifty ImageNet images from each
of forty ImageNet synsets as stimuli.
With one exception, we employed the same ImageNet synsets as classes
(Table~\ref{tab:classes}).
Since we sought 1,000 images from each class, and one class, \texttt{n03197337},
\emph{digital watch}, contained insufficient images at time of download, we
replaced that class with \texttt{n04555897}, \emph{watch}.

\begin{table}[t]
  \begin{center}
    \caption{ImageNet synsets employed as classes in our experiment.}
    \resizebox{\textwidth}{!}{\begin{tabular}{@{}lllll@{}}
      \texttt{n02106662} \emph{German shepherd} &
      \texttt{n02124075} \emph{Egyptian cat} &
      \texttt{n02281787} \emph{lycaenid} &
      \texttt{n02389026} \emph{sorrel} &
      \texttt{n02492035} \emph{capuchin} \\
      \texttt{n02504458} \emph{African elephant} &
      \texttt{n02510455} \emph{giant panda} &
      \texttt{n02607072} \emph{anemone fish} &
      \texttt{n02690373} \emph{airliner} &
      \texttt{n02906734} \emph{broom} \\
      \texttt{n02951358} \emph{canoe} &
      \texttt{n02992529} \emph{cellular telephone} &
      \texttt{n03063599} \emph{coffee mug} &
      \texttt{n03100240} \emph{convertible} &
      \texttt{n03180011} \emph{desktop computer} \\
      \texttt{n04555897} \emph{watch} &
      \texttt{n03272010} \emph{electric guitar} &
      \texttt{n03272562} \emph{electric locomotive} &
      \texttt{n03297495} \emph{espresso maker} &
      \texttt{n03376595} \emph{folding chair} \\
      \texttt{n03445777} \emph{golf ball} &
      \texttt{n03452741} \emph{grand piano} &
      \texttt{n03584829} \emph{iron} &
      \texttt{n03590841} \emph{jack-o-lantern} &
      \texttt{n03709823} \emph{mailbag} \\
      \texttt{n03773504} \emph{missile} &
      \texttt{n03775071} \emph{mitten} &
      \texttt{n03792782} \emph{mountain bike} &
      \texttt{n03792972} \emph{mountain tent} &
      \texttt{n03877472} \emph{pajama} \\
      \texttt{n03888257} \emph{parachute} &
      \texttt{n03982430} \emph{pool table} &
      \texttt{n04044716} \emph{radio telescope} &
      \texttt{n04069434} \emph{reflex camera} &
      \texttt{n04086273} \emph{revolver} \\
      \texttt{n04120489} \emph{running shoe} &
      \texttt{n07753592} \emph{banana} &
      \texttt{n07873807} \emph{pizza} &
      \texttt{n11939491} \emph{daisy} &
      \texttt{n13054560} \emph{bolete}\\
    \end{tabular}}
    \label{tab:classes}
  \end{center}
\end{table}

We downloaded all ImageNet images of each of the forty classes that were
available on 28 July 2019, randomly selected 1,000 images for each class,
resized them to 1920$\times$1080, preserving aspect ratio by padding them with
black pixels either on the left and right or top and bottom, but not both, to
center the image.
All but one such image was either RGB or grayscale.
One image, n02492035\_15739, was in the CMYK color space so was transcoded to
RGB for compatibility with our stimulus presentation software.

The 40,000 images were partitioned into 100 sets of 400 images each.
Each set of 400 images contained exactly ten images of each of the forty
classes.
Each set of 400 images was randomly permuted.
The order of the 100 sets of images was also randomly permuted.

A single adult male subject viewed all 100 sets of images while recording EEG.\@
Recording was conducted over ten sessions.
Each session nominally recorded data from ten sets of images, though some
sessions contained fewer sets, some sessions contained more sets, and some sets
were repeated due to experimenter error.
(Runs per session: 10, 8, 10, 11, 11, 10, 10, 10, 10, 10.
Run 19 was repeated after run 20 because one image was discovered to be in
CYMK.\@
Run 43 was repeated because one earlobe electrode was off.)
When sets were repeated, only one error-free set was retained.
Each recording session was nominally about six hours in duration.
The subject typically took breaks after every three or so sets of images.
As the EEG lab was being used for other experiments as well, recording was
conducted over roughly a half-year period.
(Session dates: 21, 28 Aug 2019, 3, 10, 16, 17 Sep 2019, 13, 14, 20, 21 Jan
2020.)

Our design is counterbalanced at the whole experiment level, the session level,
and the run level.
Each unit (experiment, session, or run) has the same number of stimulus
presentations for each class with no duplicates.
Thus at any level, the baseline performance is chance.
This allows partial analyses of arbitrary combinations of individual runs or
sessions with simple calculation of statistical significance.

Each set of 400 images was presented in a single EEG run lasting 20~minutes
and 20~seconds.
Each run started with 10~s of blanking, followed by 400 stimulus
presentations, each lasting 2~s, with 1~s of blanking between adjacent stimulus
presentations, followed by 10~s of blanking at the end of the run.
There was no block structure within each run.\footnote{\citep{spampinato2017}
  employed a design where stimuli were presented in blocks of fifty images.
  Each stimulus was presented for 0.5~s with no blanking between images, but
  with 10~s blanking between blocks.
  During a pilot run of our experiment with this design, the subject reported
  that it was difficult and tedious to attend to the stimuli when presented
  rapidly without pause, thus motivating adoption of our modified design.
  Our longer trials with pauses attempt to reduce the potential of
  cross-stimulus contamination.}

EEG data was recorded from 96 channels at 4,096~Hz with 24-bit resolution using
a BioSemi ActiveTwo recorder\footnote{The ActiveTwo recorder employs 64$\times$
  oversampling and a sigma-delta A/D converter, yielding less quantization
  noise than 24-bit uniform sampling.} and a BioSemi gel electrode cap.
Two additional channels were used to record the signal from the earlobes for
rereferencing.
A trigger was recorded in the EEG data to indicate stimulus onset.
Preprocessing software verifies that there are exactly 400 triggers in each
recording.\footnote{Due to experimenter error, one recording, run 14, continued
  beyond 400 stimulus presentations.
  The recordings for the extra stimulus presentations were harmlessly
  discarded.}

The current analysis uses only the first 500~ms after stimulus onset for each
stimulus presentation, even though 2~s of data were recorded.
Further, the current analysis decimated the data from 4,096~Hz to 1,024~Hz.
This was done to speed the analysis.
The full dataset is available for potential future enhanced analysis.

Each session was recorded with a single capping with the cap remaining in place
when the subject took breaks between runs.
With fMRI data, the anatomical information captured can be used to align
volumes within a run to compensate for subject motion, between runs to
compensate for subjects exiting and reentering the scanner (co-registration),
and between subjects to compensate for variations in brain anatomy (spatial
normalization).
In contrast, for EEG data, there are no established methods to adjust for
differing brain/scalp anatomy when combining data from multiple subjects; often
approximately corresponding scalp locations are treated as equivalent.
For this reason, we recorded data from a single subject to eliminate the need
to align across subjects.
By performing capping only once per session and choosing a cap size to yield a
snug fit, any within-session alignment issues are obviated.
To minimize across-session misalignment, the same cap with pre-cut ear holes
was used across sessions with the vertex marking on the cap (location Cz)
positioned to be geodesically equidistant from the the nasion and inion in the
front-back direction, and equidistant from the left and right pre-auricular
points in the left-right direction.
Furthermore, visual inspection was done from vantage points directly in front
and at the back of the subject to check that the FPz, Fz, Cz, Pz, and Oz
markings on the cap fell on the geodesic connecting the nasion and inion.

To check whether the subject consistently viewed the images presented, online
trial averaging of the EEG data was performed in every session to obtain evoked
responses that are phase-locked to the onset of the images.
Data from two occipital channels (C31 and C32) were bandpass filtered in the
1--40~Hz range and epochs of 800~ms duration were segmented out synchronously
following the onset of each image.
Epochs with peak-to-trough fluctuations exceeding 100~mV were discarded and the
remaining epochs were averaged together to yield an 800~ms-long evoked
response.
A clear and robust N1-P2 onset response pattern was discernible in the evoked
response traces obtained in each of the 100 runs, consistent with the subject
viewing the images as instructed.
Note that all online averaging procedures (\eg\ filtering) were done to data
in a separate buffer; the raw unprocessed data from 96~channels was saved for
offline analysis.

\section{Preprocessing}

The raw EEG data was recorded in bdf file format, a single file for each of the
100 runs.\footnote{We will release the raw data and all code discussed in this
  manuscript upon publication.}
We performed minimal preprocessing on this data, independently for each run,
first rereferencing the data to the earlobes, then extracting exactly 0.5~s of
data starting at each trigger, then z-scoring each channel of the extracted
samples for each run independently, so that the extracted samples for each
channel of each run have zero mean and unit variance, and then finally
decimating the signal from 4,096~Hz to 1,024~Hz.
No filtering was performed.
After rereferencing, there is no appreciable line noise to filter.
Randomized trials preclude classifying long-term transients, thus there is no
need to filter out such transients.
Note that this preprocessing is minimal; future studies should consider
improving the SNR of the neural signals by manually removing artifacts from eye
blinks, movements, and facial muscle artifacts.

The data was then randomly partitioned into five equal-sized folds, each
containing the same number of samples of each class.
All analyses below report average across five-fold round-robin
leave-one-fold-out cross validation, taking four folds in each split as
training data and the remaining fold as test data.
When performing analyses on subsets of the data, the sizes of the folds, and
thus the sizes of the training and test sets, varied proportionally.

\section{Classifiers}

The analyses below employ five different classifiers, an LSTM
\citep{hochreiter1997}, a nearest neighbor classifier ($k$-NN), an SVM,
a two-layer fully-connected neural network (MLP), and a one-dimensional CNN.\@
The LSTM is the same as Spampinato \etal\citep{spampinato2017} with the
modifications discussed previously by us \citep{li2020}.
The remainder are as described previously by us \citep{li2020}, with minor
differences resulting from the fact that here the signals contain 512 temporal
samples instead of 440.

\section{Analyses}

To answer the first question, \emph{Is it possible to decode object
class from EEG data recorded from subjects viewing image stimuli with
randomized stimulus presentation order?}, we trained and tested each of the
five classifiers on the entire dataset of 1,000 stimulus presentations of each
of forty classes, using five-fold cross validation (Table~\ref{tab:accuracy}).
All analyses here and below test statistical significance above chance using
$p<\textrm{0.005}$ against a null hypothesis by a binomial cmf.
Only two classifiers, the SVM and the 1D~CNN, yield statistically significant
above-chance accuracy.

\begin{table}[t]
  \begin{center}
    \caption{Classification accuracy on the validation set,
      averaged across all five folds, for each classifier.
      Here and throughout, starred values indicate statistical significance
      above chance ($p<\textrm{0.005}$) by a binomial cmf.}
    % pkl.load(open("/home/qobi/eeg-experiments/imagenet40-1000/analysis/run-100/MLP.pkl", "rb"))
    % 0.02465000133961439, 0.6773809339093255
    % tail -n 1 ~/eeg-experiments/imagenet40-1000/analysis/run-100/stat-image-MLP.txt
    % 0.0237500010058      0.9479963463445544
    % So differs from last row of Table 3 (left).
    % Likely also LSTM, K-NN, SVM, and CNN but these haven't finished yet so
    % can't check.
    % CNN agrees with last row of Table 3 (right). But can't check SVM yet
    % because hasn't finished yet.
    % Need to check consistency between last row of Table 3 (left) and last row
    % of Table 3 (right).
    \input{image-full-accuracy}
    \label{tab:accuracy}
  \end{center}
\end{table}

\begin{figure}[t]
  \centering
  \includegraphics[width=0.7\textwidth]{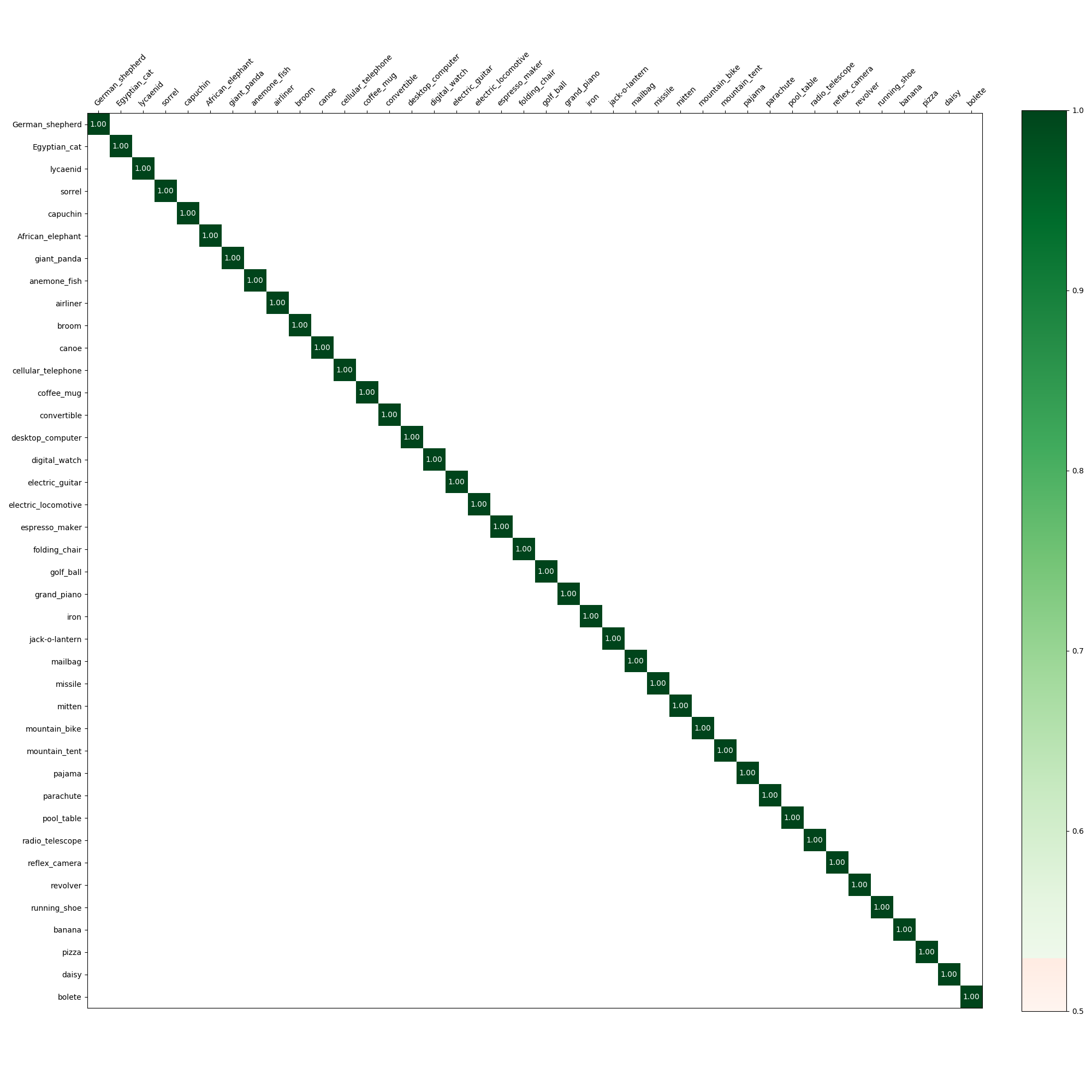}
  \caption{Two-class classification accuracy on the validation set, averaged
    across all five folds, for each pair of classes and the LSTM classifier.
    Green denotes statistical significance above chance ($p<\textrm{0.005}$) by
    a binomial cmf.
    Red denotes above chance but not statistically significant.
    Blank denotes at or below chance.}
  \label{fig:pairwise-LSTM}
\end{figure}

\begin{figure}[t]
  \centering
  \includegraphics[width=0.7\textwidth]{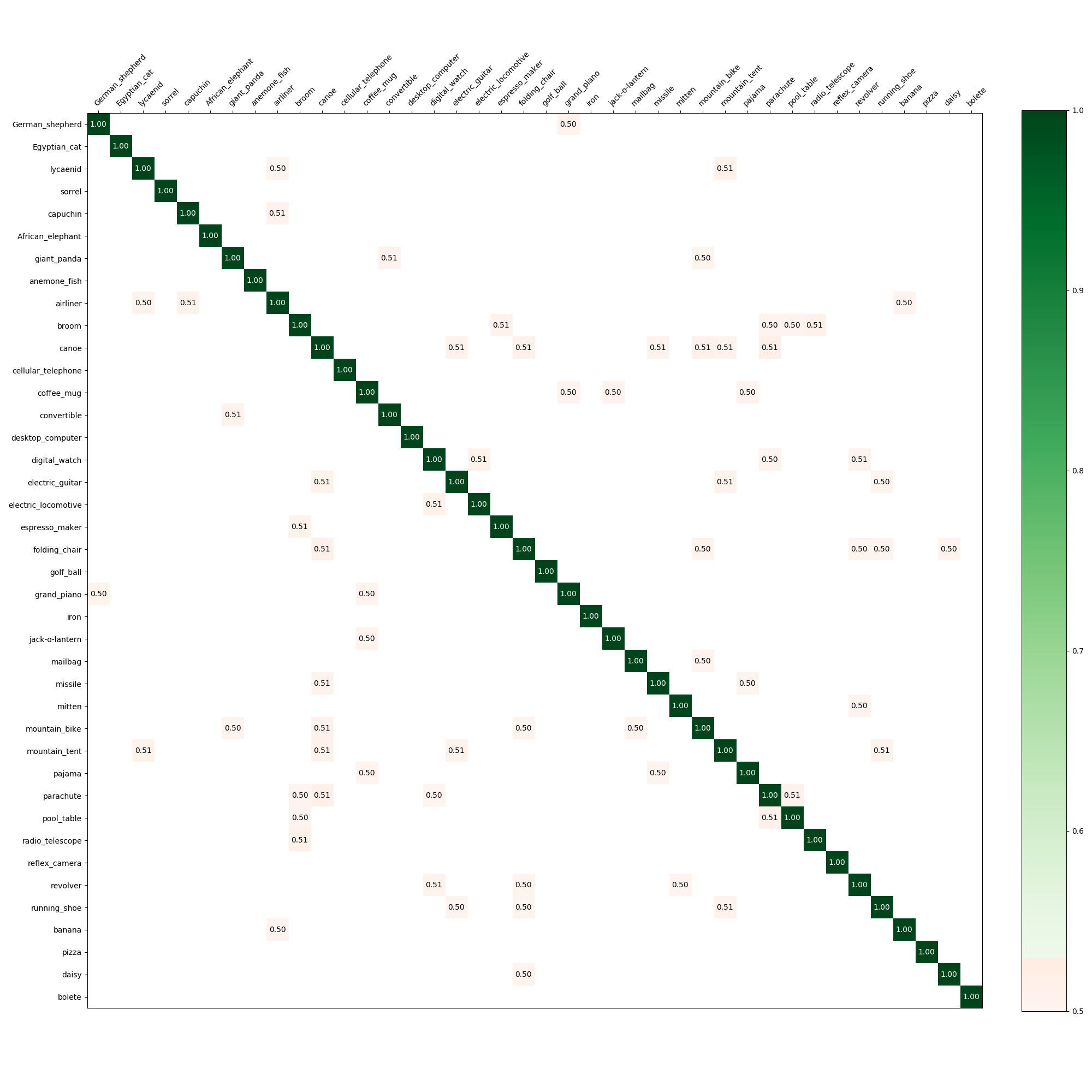}
  \caption{Variant of Fig.~\protect\ref{fig:pairwise-LSTM} for the $k$-NN
    classifier.}
  \label{fig:pairwise-K-NN}
\end{figure}

\begin{figure}[t]
  \centering
  \includegraphics[width=0.7\textwidth]{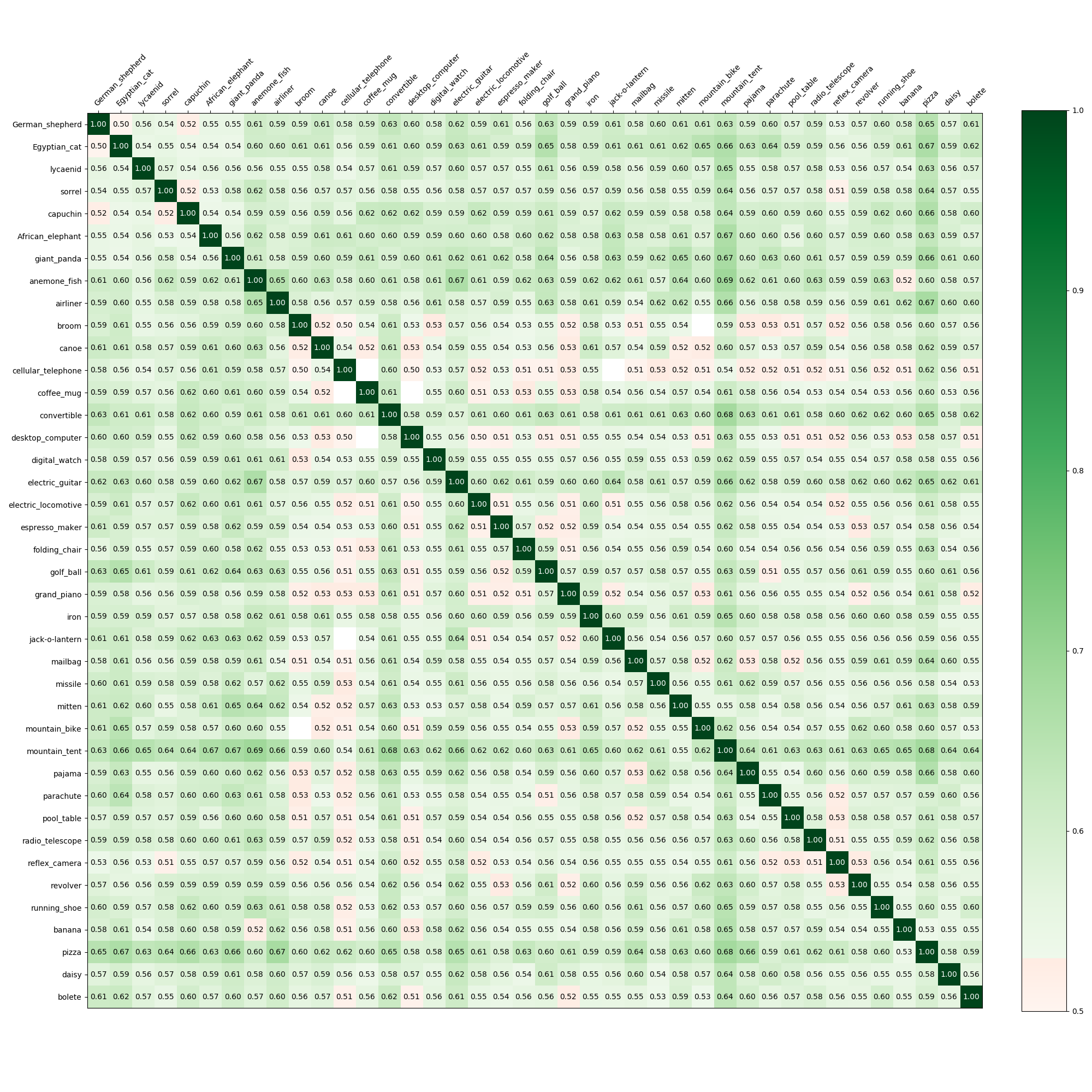}
  \caption{Variant of Fig.~\protect\ref{fig:pairwise-LSTM} for the SVM
    classifier.}
  \label{fig:pairwise-SVM}
\end{figure}

\begin{figure}[t]
  \centering
  \includegraphics[width=0.7\textwidth]{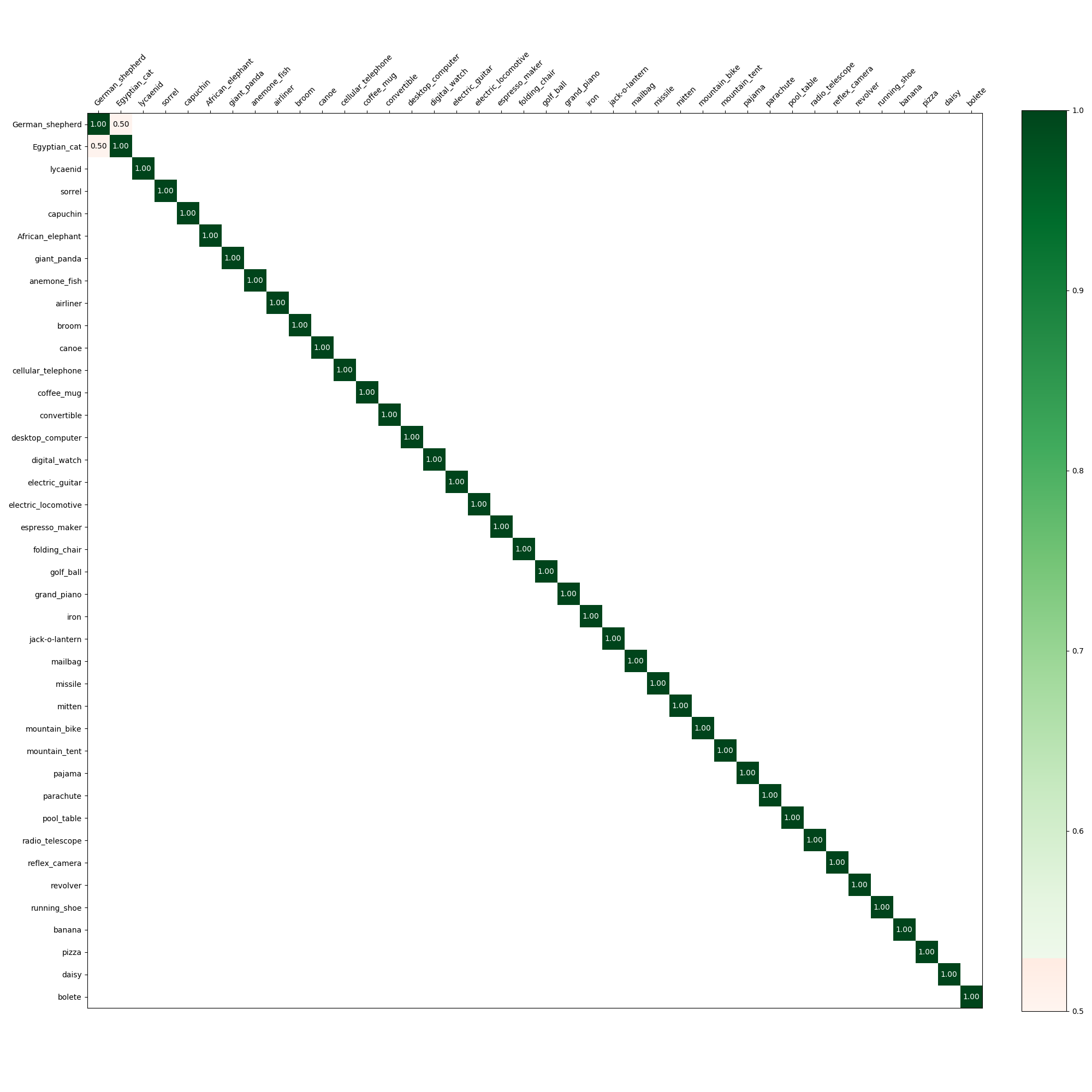}
  \caption{Variant of Fig.~\protect\ref{fig:pairwise-LSTM} for the MLP
    classifier.}
  \label{fig:pairwise-MLP}
\end{figure}

\begin{figure}[t]
  \centering
  \includegraphics[width=0.7\textwidth]{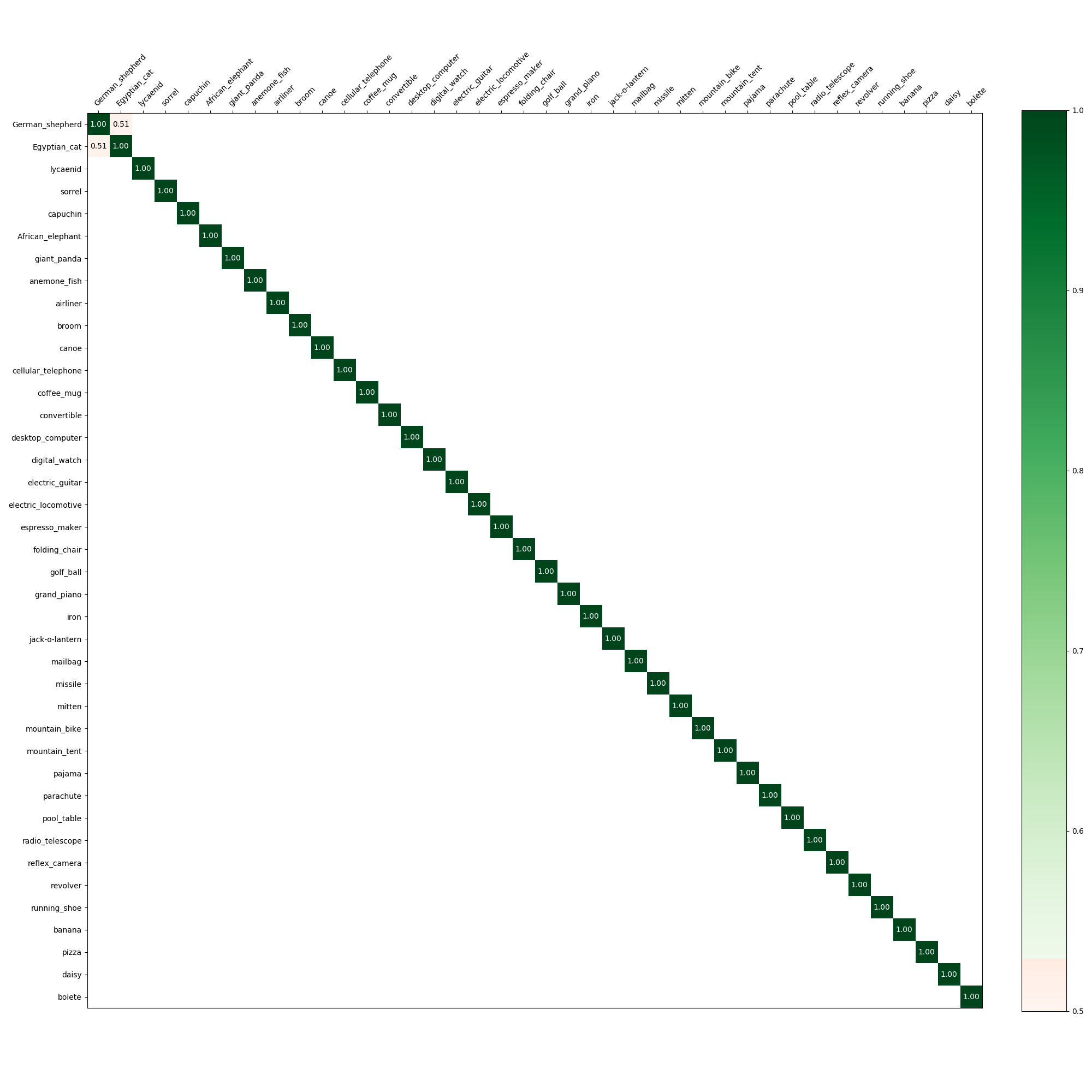}
  \caption{Variant of Fig.~\protect\ref{fig:pairwise-LSTM} for the 1D~CNN
    classifier.}
  \label{fig:pairwise-CNN}
\end{figure}

\begin{figure}[t]
  \centering
  \resizebox{\textwidth}{!}{\begin{tabular}{@{}cc@{}}
    \includegraphics{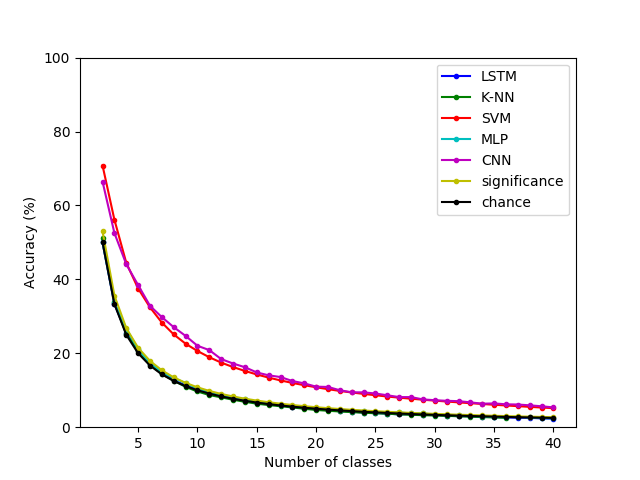}&
    \includegraphics{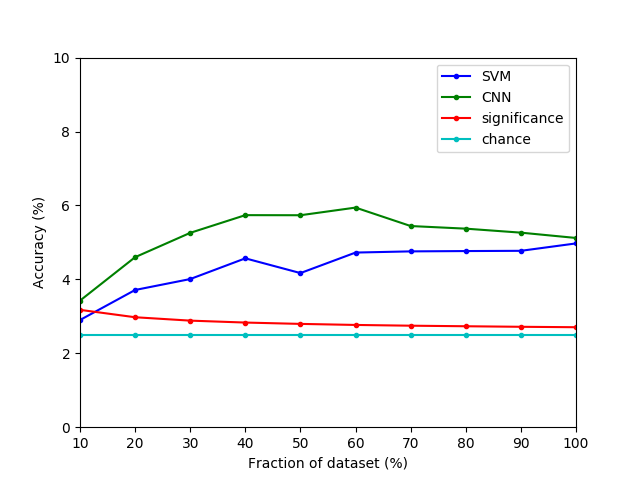}\\
  \end{tabular}}
  \caption{(left)~Classification accuracy on the validation set, averaged
    across all five folds, as a function of the number of classes, for each
    classifier, for the most discriminable subset of classes as determined
    by the greedy algorithm.
    This plot contains all data points for the LSTM, SVM, MLP, and 1D~CNN
    classifiers and 35~data points for the $k$-NN classifier.
    The remaining classifier runs are in progress.
    (They require several months on our cluster of 54 GPUs.)
    This plot will be replaced with a complete plot in a final version of this
    manuscript if accepted.
    (right)~Classification accuracy on the validation set, for all forty
    classes as a function of the fraction of the dataset used for train and
    test, for the two classifiers for which accuracy is above chance in a
    statistically significant fashion.
    Significance denotes above chance ($p<\textrm{0.005}$) by a binomial cmf.
    Tabular versions of these plots are in Table~\ref{tab:add}.}
  \label{fig:add}
\end{figure}

To answer the second question, \emph{How many distinct classes can one decode?},
we performed a greedy analysis, independently for each classifier.
We first trained and tested a classifier for each pair of distinct classes.
Figs.~\ref{fig:pairwise-LSTM}--\ref{fig:pairwise-CNN} depict the resulting
average validation accuracies.
Only one classifier, the SVM, yielded a statistically significant above-chance
accuracy for some pair.
It did so for a large number of pairs.
We then selected the pair with the highest average validation accuracy,
independently for each classifier, and selected the first element of this pair
as the seed for a class sequence for that classifier.
Then for each~$n$ between two and forty, we greedily and incrementally added one
more class to the class sequence for each classifier.
This class was selected by trying each unused class, adding it to the class
sequence, training and testing a classifier with that addition, and selecting
the added class that led to the highest classification accuracy.
This yielded a distinct class sequence of next-most-discriminable classes for
each classifier, along with an average validation accuracy on each initial
prefix of that sequence (Fig.~\ref{fig:add} left and Table~\ref{tab:add} left).
With the exception of a single data point, the MLP classifier achieving
marginally significant above-chance classification accuracy for $n=29$, only
two classifiers, the SVM and the 1D~CNN, yielded statistically significant
above-chance accuracy for any number of classes.
They both yielded statistically significant above-chance accuracy for all
numbers of classes.

\begin{table}[t]
  \begin{center}
    \caption{Tabular version of Fig.~\ref{fig:add}.}
    \resizebox{0.8\textwidth}{!}{\begin{tabular}{@{}c@{\hspace*{20pt}}c@{}}
      \begin{tabular}{@{}p{234pt}@{}}
        \input{image-accuracy}
      \end{tabular}&
      \raisebox{163pt}{\begin{tabular}{@{}p{179pt}@{}}
        \input{image-partial-accuracy}
      \end{tabular}}\\
    \end{tabular}}
    \label{tab:add}
  \end{center}
\end{table}

To answer the third question, \emph{How much training data is needed?},
we performed an analysis where classifiers were trained and tested on
progressively larger portions of the dataset, starting with 10\%, incrementing
by 10\%, until the full dataset was tested.
This was done by taking the first ten runs and incrementally adding the next
ten runs.
This was done only for the SVM and the 1D~CNN, as only these had statistically
significant above-chance accuracy (Fig.~\ref{fig:add} right and
Table~\ref{tab:add} right).
Validation accuracy generally increases with the availability of more training
data, though growth tapers off demonstrating diminishing returns.

Finally, the fourth question, \emph{What classification architectures allow
  such decoding?}, was implicitly answered by the above three analyses.
Only the SVM and the 1D~CNN answer any of the above three questions in the
affirmative.
The SVM and the 1D~CNN answer all of the above three questions in the
affirmative.

\section{Significance}

With our data collection, each run lasted 20:20.
The recording alone for each session nominally took 3:23:20.
Including capping, uncapping, subject breaks, setup, teardown, and data
transfer, each session took more than six hours, \ie\ most of a full business
day.
The ten sessions required to collect our dataset took more than sixty hours,
\ie\ most of two full business weeks.
Few subjects would consent to, and complete, such an extensive and tedious data
collection effort.
Consider what it would take to collect a larger dataset.
Collecting EEG recordings of a single subject viewing all 1,431,167 images of
ILSVRC 2012 \citep{russakovsky2014} would take more than a full business year
with the protocol employed in this manuscript.
Doing so for all 14,197,122 images and 21,841 synsets currently included in
ImageNet (3 Feb 2020) would take more than a full business decade.
We doubt that any subject would consent to, and complete, such an extensive and
tedious data collection effort.
Moreover, we doubt that any EEG lab would dedicate the resources needed to do
so.

\section{Related Work}

We know of two prior attempts at collecting large EEG datasets.
% (+ 67635 (/ 910476 14) (/ 163932 4) (/ 65250 5))
The ``MNIST'' of Brain Digits recorded EEG data from a single subject viewing
186,702 presentations of the digits 0--9, each for 2~s, over a two-year period
\citep{vivancos2019}.
(While this dataset is called ``MNIST,'' it is unclear what stimuli the subject
viewed.)
% mindwave 1 https://www.mdpi.com/1424-8220/19/12/2808/htm
% epoc 16
% https://www.researchgate.net/figure/The-positions-of-16-electrodes-of-the-Emotiv-EPOC-headset_fig3_236739621
% muse 5, one reference https://www.frontiersin.org/articles/10.3389/fnins.2017.00109/full
% insight 5 https://www.emotiv.com/insight/
It was recorded by the subject themselves with four different consumer-grade
EEG recording devices (Neurosky Mindwave, Emotiv EPOC, Interaxon Muse, and
Emotiv Insight), each with only a handful of electrodes (Mindwave: 1, EPOC: 14,
Muse: 4, and Insight: 5).
``IMAGENET'' of The Brain recorded EEG data from a single subject viewing
14,012 stimulus presentations spanning 13,998 ILSVRC 2013 training images and
569 classes, each for 3~s, over a one-year period \citep{vivancos2018}.
The number of images per class ranged from 8 to 44.
Fourteen images were presented as stimuli twice.
It was recorded by the subject themselves with a single consumer-grade EEG
recording device (Emotiv Insight) with five electrodes.
(The number of `brain signals' reported by Vivancos \citep{vivancos2019,
  vivancos2018} differ from the above due to multiplication of the stimulus
presentations by the number of electrodes.)

While we applaud such efforts, several issues arise with these datasets.
Consumer grade recording devices have far fewer electrodes, far lower sample
rate, and far lower resolution than research-grade EEG recording devices.
They use dry electrodes instead of gel electrodes.
There is no control over electrode placement.
It is unclear how to use recordings from different devices with different
numbers and configurations of electrodes as part of a common experiment.
The designs were not counterbalanced.
The stimulus presentation order is not clear so it is not clear whether these
datasets suffer from the issues described previously by us \citep{li2020}.
The recording did not appear to employ a trigger so it is unclear how to
determine the stimulus onset.
The reduced precision limits the utility of these datasets.
Moreover, the ``MNIST'' of Brain Digits has too few classes and ``IMAGENET'' of
The Brain has too few stimuli per class to answer the questions we pose here.

A significant amount of prior work suffers irreparably from flawed EEG
experiment design.
The dataset collected by Spampinato \etal\citep{spampinato2017} is contaminated
by its combination of block design and having all images of a class appear in
only one block.
Unfortunately, this fundamental design confound cannot be overcome by post
processing.
Considerable follow-on work \citep{palazzo2017, kavasidis2017, palazzo2018,
  palazzo2018recent, du2018multi, Bozal2017, jiang2019context, jiaodecoding,
  zhang2018using, du2019doubly, zhong2018elstm, mukherjee2019cogni,
  fares2018region, hwang2019ezsl} that uses this dataset also inherits this
confound and their conclusions may thus be flawed.
We previously \citep{li2020} demonstrated that accuracy drops to chance when
such flawed designs are replaced with randomized trials keeping all other
aspects of the experiment design unchanged, including the dataset size.
Here we demonstrate that accuracy increases to only marginally above
significance even when the dataset size is increased to the bounds of
feasibility.

\section{Conclusion and Summary of Novel Contributions}

In this manuscript we demonstrate five novel contributions.
\begin{compactenum}
\item We show that it does not seem possible to decode object class from EEG
  data recorded from subjects viewing image stimuli with randomized stimulus
  presentation order when the dataset contains between two and forty classes
  with classification accuracy that is above chance in a statistically
  significant fashion using an LSTM (the classifier employed by Spampinato
  \etal\citep{spampinato2017}), a $k$-NN, or an MLP classifier, even if one has
  a training set that is 20$\times$ larger than previous work.
  It appears that LSTM, $k$-NN, and MLP classifiers are ill-suited to
  classifying object class from EEG data recorded from subjects viewing image
  stimuli no matter how many classes are classified and no matter how much
  training data is available.
  This refutes a large amount of prior work and shows that the task attempted
  by that work is simply infeasible.
\item We show that it is possible to decode object class from EEG data
  recorded from subjects viewing image stimuli with randomized stimulus
  presentation order when the dataset contains between two and forty classes
  with classification accuracy that is marginally above chance in a
  statistically significant fashion using either an SVM classifier or the
  1D~CNN classifier proposed previously by us \citep{li2020}.
  However, it is not possible to obtain accuracy above chance in a
  statistically significant fashion with a dataset of the size employed by
  previous work (fifty samples per class).
  For forty classes, accuracy is marginally below statistical significance
  for the SVM and marginally above statistical significance for the 1D~CNN with
  100 samples per class (2$\times$ previous work) and increases to about 5\%
  for the SVM and 6\% for the 1D~CNN with about 600 samples per class
  (12$\times$ previous work) and then tapers off.
  It appears that no amount of additional training data will allow
  substantially better classification accuracy for forty classes using the
  classifiers that we tried.
\item Our classification accuracies are state-of-the-art for decoding object
  class from EEG data recorded from subjects viewing image stimuli with
  randomized stimulus presentation and large numbers of classes.
  To our knowledge, these are also the first results yielding statistically
  significant above-chance accuracy with a large number of classes.
  Previous reports of higher accuracy, to the best of our knowledge, appear
  to use data that are contaminated by the confounds we describe.
\item We show that gathering the amounts of training data to achieve this
  level of accuracy are at the bounds of feasibility.
  Gathering the requisite data to train classifiers for a larger number of
  classes, such as all of ILSVRC 2012, let alone all of ImageNet, would require
  Herculean effort.
\item We collected by far the largest known dataset of EEG recordings from a
  single subject viewing image stimuli with professional grade equipment and
  procedures using proper randomized trials.
  It has 20$\times$ as many stimuli per class as our previous dataset
  \citep{li2020}, 4$\times$ as many classes as the dataset of Vivancos
  \citep{vivancos2019} (which is not known to have randomized trials), and
  23$\times$ to 125$\times$ as many stimuli per class as the dataset of
  Vivancos \citep{vivancos2018} (which is also not known to have randomized
  trials).
  We will release this dataset upon publication.
  This will facilitate experimentation with new classification and analysis
  methods that will hopefully lead to improved accuracy in the future.
\end{compactenum}
Despite recent claims to the contrary, presented to the computer-vision
community with great fanfare, the problem of classifying visually perceived
objects from EEG recordings with high accuracy for large numbers of classes is
immensely difficult and currently beyond the state of the art.
It appears to be infeasible and may even be impossible.
A common euphoria in the community is that large datasets have allowed
deep-learning methods to solve practically everything.
It appears, however, to have reached its limit with object classification from
EEG recordings.
Neither heroic amounts of data, at the bounds of feasibility, nor the standard
deep-learning architectures of fully connected networks (MLP), convolutional
neural networks (CNN), or recurrent neural networks (LSTM)---or even more
traditional machine-learning methods like nearest-neighbor classifiers ($k$-NN)
or support vector machines (SVM), appear suited to the task.
We present our data and this task to the community as a challenge problem for
moving beyond large datasets and deep learning, to true understanding of human
visual perception.

\ifnum\value{submission}=1
\else
\section*{Acknowledgments}

This work was supported, in part, by the US National Science Foundation under
Grants 1522954-IIS and 1734938-IIS, by the Intelligence Advanced Research
Projects Activity (IARPA) via Department of Interior/Interior Business Center
(DOI/IBC) contract number D17PC00341, and by Siemens Corporation, Corporate
Technology.
Any opinions, findings, views, and conclusions or recommendations expressed in
this material are those of the authors and do not necessarily reflect the
views, official policies, or endorsements, either expressed or implied, of the
sponsors.
The U.S. Government is authorized to reproduce and distribute reprints for
Government purposes, notwithstanding any copyright notation herein.
\fi

\bibliographystyle{splncs04}
\bibliography{eccv2020a}
\end{document}

%% file: image-full-accuracy.tex
\begin{tabular}{@{}rrrrr@{}}
LSTM&$k$-NN&SVM&MLP&1D CNN\\
\hline
2.2\%\notsig&2.1\%\notsig&5.0\%\sig&2.5\%\notsig&5.1\%\sig\\
\end{tabular}

%% file: image-accuracy.tex
\begin{tabular}{@{}r|rrrrr@{}}
&\multicolumn{5}{c}{accuracy}\\
number of classes&LSTM&$k$-NN&SVM&MLP&1D CNN\\
\hline
2&50.0\%\notsig&51.3\%\notsig&70.8\%\sig&50.0\%\notsig&66.4\%\sig\\
3&33.3\%\notsig&33.8\%\notsig&56.1\%\sig&33.7\%\notsig&52.5\%\sig\\
4&25.5\%\notsig&25.1\%\notsig&44.5\%\sig&26.7\%\notsig&44.1\%\sig\\
5&20.8\%\notsig&20.7\%\notsig&37.5\%\sig&21.1\%\notsig&38.4\%\sig\\
6&17.1\%\notsig&16.9\%\notsig&32.4\%\sig&17.4\%\notsig&32.8\%\sig\\
7&14.8\%\notsig&14.4\%\notsig&28.3\%\sig&14.9\%\notsig&29.8\%\sig\\
8&12.7\%\notsig&12.6\%\notsig&25.1\%\sig&13.3\%\notsig&27.1\%\sig\\
9&11.3\%\notsig&10.9\%\notsig&22.6\%\sig&11.9\%\notsig&24.7\%\sig\\
10&10.1\%\notsig&9.7\%\notsig&20.6\%\sig&10.5\%\notsig&22.0\%\sig\\
11&9.4\%\notsig&8.7\%\notsig&18.9\%\sig&9.2\%\notsig&20.9\%\sig\\
12&8.4\%\notsig&8.1\%\notsig&17.5\%\sig&8.7\%\notsig&18.4\%\sig\\
13&8.0\%\notsig&7.4\%\notsig&16.3\%\sig&8.2\%\notsig&17.2\%\sig\\
14&7.2\%\notsig&6.9\%\notsig&15.2\%\sig&7.5\%\notsig&16.2\%\sig\\
15&6.7\%\notsig&6.4\%\notsig&14.3\%\sig&6.9\%\notsig&14.8\%\sig\\
16&6.2\%\notsig&6.0\%\notsig&13.4\%\sig&6.5\%\notsig&14.0\%\sig\\
17&5.9\%\notsig&5.7\%\notsig&12.7\%\sig&6.1\%\notsig&13.6\%\sig\\
18&5.5\%\notsig&5.3\%\notsig&12.0\%\sig&5.8\%\notsig&12.5\%\sig\\
19&5.1\%\notsig&5.0\%\notsig&11.4\%\sig&5.4\%\notsig&11.9\%\sig\\
20&4.8\%\notsig&4.7\%\notsig&10.8\%\sig&5.3\%\notsig&11.0\%\sig\\
21&4.7\%\notsig&4.5\%\notsig&10.3\%\sig&4.9\%\notsig&10.9\%\sig\\
22&4.4\%\notsig&4.2\%\notsig&9.8\%\sig&4.8\%\notsig&10.0\%\sig\\
23&4.2\%\notsig&4.1\%\notsig&9.4\%\sig&4.5\%\notsig&9.4\%\sig\\
24&4.0\%\notsig&3.9\%\notsig&9.0\%\sig&4.4\%\notsig&9.4\%\sig\\
25&3.8\%\notsig&3.8\%\notsig&8.6\%\sig&4.0\%\notsig&9.1\%\sig\\
26&3.7\%\notsig&3.6\%\notsig&8.3\%\sig&3.9\%\notsig&8.6\%\sig\\
27&3.5\%\notsig&3.5\%\notsig&8.0\%\sig&4.0\%\notsig&8.1\%\sig\\
28&3.5\%\notsig&3.4\%\notsig&7.7\%\sig&3.7\%\notsig&8.1\%\sig\\
29&3.3\%\notsig&3.3\%\notsig&7.4\%\sig&3.8\%\sig&7.5\%\sig\\
30&3.4\%\notsig&3.2\%\notsig&7.2\%\sig&3.3\%\notsig&7.3\%\sig\\
31&3.1\%\notsig&3.1\%\notsig&6.9\%\sig&3.4\%\notsig&7.1\%\sig\\
32&3.0\%\notsig&3.0\%\notsig&6.7\%\sig&3.2\%\notsig&7.0\%\sig\\
33&3.0\%\notsig&2.8\%\notsig&6.5\%\sig&3.1\%\notsig&6.7\%\sig\\
34&2.8\%\notsig&2.7\%\notsig&6.3\%\sig&2.9\%\notsig&6.4\%\sig\\
35&2.8\%\notsig&2.6\%\notsig&6.1\%\sig&2.8\%\notsig&6.4\%\sig\\
36&2.6\%\notsig&2.5\%\notsig&5.9\%\sig&2.7\%\notsig&6.2\%\sig\\
37&2.6\%\notsig&\notsig&5.7\%\sig&2.8\%\notsig&6.1\%\sig\\
38&2.5\%\notsig&\notsig&5.5\%\sig&2.6\%\notsig&5.9\%\sig\\
39&2.4\%\notsig&\notsig&5.3\%\sig&2.5\%\notsig&5.7\%\sig\\
40&2.3\%\notsig&\notsig&5.2\%\sig&2.4\%\notsig&5.4\%\sig\\
\end{tabular}

%% file: image-partial-accuracy.tex
\begin{tabular}{@{}r|rr@{}}
&\multicolumn{2}{c}{accuracy}\\
fraction of dataset&SVM&1D CNN\\
\hline
10\%&2.9\%\notsig&3.4\%\sig\\
20\%&3.7\%\sig&4.6\%\sig\\
30\%&4.0\%\sig&5.3\%\sig\\
40\%&4.6\%\sig&5.7\%\sig\\
50\%&4.2\%\sig&5.7\%\sig\\
60\%&4.7\%\sig&5.9\%\sig\\
70\%&4.8\%\sig&5.4\%\sig\\
80\%&4.8\%\sig&5.4\%\sig\\
90\%&4.8\%\sig&5.3\%\sig\\
100\%&5.0\%\sig&5.1\%\sig\\
\end{tabular}